\documentclass[usenatbib]{mn2e}
\usepackage{epsfig}
\usepackage{subfigure}
\usepackage{amssymb}

\newcommand{\bn}{\begin{enumerate}}
\newcommand{\en}{\end{enumerate}}
\newcommand{\bi}{\begin{itemize}}
\newcommand{\ei}{\end{itemize}}
\newcommand{\rhoth}{\rho_{\rm th}}
\newcommand{\Nmin}{N_{\rm min}}
\newcommand{\Zsun}{Z_\odot}
\newcommand{\Msun}{M_\odot}
\newcommand{\Mstar}{M_{\rm star}}
\newcommand{\Mgas}{M_{\rm gas}}
\newcommand{\overd}{\rho/\bar{\rho}}
\newcommand{\fgas}{f_{\rm gas}}

\newcommand{\HI}{H{\sc i}\,\,}
\newcommand{\NHI}{{N_{\rm HI}}}

\newcommand{\apj}{ApJ}
\newcommand{\aap}{A\&A}
\newcommand{\apjl}{ApJL}
\newcommand{\apjs}{ApJS}
\newcommand{\mnras}{MNRAS}

\newcommand{\araa}{ARA\&A}

\newcommand{\nat}{Nature}

\newcommand{\gca}{GeCoA}

\title{Effects of metal enrichment and metal cooling in galaxy growth and cosmic star formation history}
\author[Choi \& Nagamine]
{Jun-Hwan Choi\thanks{Email: jhchoi@physics.unlv.edu}, 
Kentaro Nagamine\thanks{Visiting Researcher, Institute for the Physics and Mathematics of the Universe, University of Tokyo, 5-1-5 Kashiwanoha Kashiwa-shi, Chiba 277-8582 Japan} \vspace{0.3cm}\\
Department of Physics \& Astronomy, University of Nevada, Las Vegas, 4505 S. Maryland Pkwy, Las Vegas, NV, 89154-4002, U.S.A.}

\begin{document}

\maketitle

\begin{abstract}
We present the results of a numerical study on the effects of metal enrichment 
and metal cooling on galaxy formation and cosmic star formation (SF) history 
using cosmological hydrodynamic simulations.  We find following differences 
in the simulation with metal cooling when compared to the run without it:
(1) the cosmic star formation rate (SFR) is enhanced by about 50 \& 20\% at 
$z=1$ \& 3, respectively; 
(2) the gas mass fraction in galaxies is lower; 
(3) the total baryonic mass function (gas + star) at $z=3$ does not differ 
significantly, but shows an increase in the number of relatively massive 
galaxies at $z=1$; 
(4) the baryonic mass fraction of intergalactic medium (IGM) is reduced
at $z<3$ due to more efficient cooling and gas accretion onto galaxies. 
Our results suggest that the metal cooling enhances the galaxy growth by 
two different mechanisms: (1) increase of SF efficiency in the local 
interstellar medium (ISM), and (2) increase of IGM accretion onto galaxies. 
The former process is effective throughout most of the cosmic history, 
while the latter is effective only at $z<3$ when the IGM is sufficiently 
enriched by metals owing to feedback. 
\end{abstract}



\section{Introduction}
\label{sec:intro}

Gas cooling plays a key role in galaxy formation.  According to the 
hierarchical structure formation model, galaxies form in gravitationally 
collapsed dark matter halos.  The dissipative baryonic matter cools by 
emitting radiation and condenses into halos to form galaxies 
\citep{Rees.Ostriker:77, White.Rees:78}.
Subsequently stars form through the radiative cooling of interstellar medium 
\citep[ISM; e.g.,][]{McKee.Ostriker:07}.

The cooling rate of gas varies with its chemical abundance, because different 
atoms and molecules have different cooling rates.  The chemical abundance of 
cosmic gas evolves as a function of time and environment.
In the early universe, the primordial gas consists of mainly hydrogen 
(mass fraction of $X\simeq 0.74$) and helium (mass fraction of $Y\simeq 0.25$) 
with a negligible amount of light metals \citep[e.g.][]{Burles.etal:01}.
Throughout this paper, all the elements heavier than helium are collectively 
called `metals'.  Most metals are produced inside stars and spread out to the 
space by galactic winds driven by supernova (SN) explosions.  
In the metal-enriched gas, the total cooling rate is significantly enhanced 
relative to that of the primordial gas by the atomic emission lines from 
recombination of ionised metals. 

Metals may influence galaxy formation in two different ways. 
Firstly, the SF efficiency is increased owing to the shorter gas cooling time. 
Star formation generally takes place in dusty metal-enriched ISM, therefore, 
the enhanced cooling rate by metals would boost up the SF efficiency. 
However, the energy/momentum feedback by SNe may suppress the subsequent star 
formation after the initial starburst by heating up the ambient gas.  
This complicates the situation, and it is not clear whether the net effect 
of metal cooling and SN feedback would be negative or positive.  
The feedback process is highly nonlinear, therefore, a direct numerical 
simulation would be a useful tool to explore the effects of feedback and 
metal cooling. 

Secondly, the metals dispersed into the IGM by SN feedback also enhance the 
cooling of IGM.  This can lead to the increase of IGM accretion onto 
galaxies, because colder gas can sink into galaxies more easily.  
Both of the above two effects may affect the galaxy growth and 
cosmic SF history significantly.  Therefore it is essential to include the 
effects of metal cooling and chemical enrichment by feedback in the studies 
of galaxy formation and evolution, and capture the complex two-way 
interactions between galaxies and IGM. 
In addition, metal enrichment also affects the equation of state by 
altering the mean molecular weight of gas, which influences the hydrodynamic 
calculation of gas thermal state. 

Cosmological hydrodynamic simulations are widely used in the studies of galaxy 
formation and cosmic SF history \citep[e.g.,][]{Cen:92, Katz.etal:92, 
Cen.Ostriker:93, Katz.etal:96, Nagamine.etal:00, Pearce.etal:01, 
Nagamine.etal:01a, Nagamine.etal:01b, 
Ascasibar.etal:02, Murali.etal:02, Nagamine:02, Weinberg.etal:02, 
Kawata.Gibson:03, Kravtsov:03, Marri.White:03, Springel.Hernquist:03:S, 
Governato.etal:04, Robertson.etal:04, Weinberg.etal:04, 
Nagamine.etal:04, Nagamine.etal:05a, Nagamine.etal:05b, Cen.etal:05, 
Keres.etal:05, Finlator.etal:06, Hoeft.etal:06, Nagamine.etal:06, 
Scannapieco.etal:06, Tasker.Bryan:06, Dave.Oppenheimer:07, Finlator.etal:07, 
Governato.etal:07, Kobayashi.etal:07, Tornatore.etal:07, Finlator.etal:08}.
Most of these works included the treatments of radiative cooling by H \& He, 
star formation, and SN feedback. However, some of the simulations did not 
include the effects of metal cooling and/or chemical enrichment by 
galactic wind. 
Furthermore, the effects of metal cooling on galaxy growth and cosmic SF 
history has not been explored in detail using cosmological hydrodynamic 
simulations and presented in the literature, as it is costly to run large 
cosmological simulations with and without the effect of metal cooling. 

In this paper, we investigate the effects of metal cooling and metal 
enrichment on galaxy growth and cosmic SF history using a series of 
cosmological hydrodynamic simulations with and without metal cooling. 
The aim of this paper is to single out the effects of metal cooling among 
our simulations.  This work can be regarded as our initial step towards the 
long-term goal of developing more complete cosmological hydrodynamic code 
with physically motivated models of star formation and feedback. 

The paper is organised as follows. 
In \S~\ref{sec:method}, we describe our simulation method focusing on how we 
implement the metal cooling. In \S~\ref{sec:global}, we study the 
metal cooling effects on the global properties, such as cosmic SFR, phase 
space ($\rho$ vs. $T$) distribution of gas, and the evolution of four 
phases (hot, warm-hot, diffuse, and condensed) of baryons.
We then study the galaxy mass functions (\S~\ref{sec:MF}) and 
gas mass fractions (\S~\ref{sec:GasFrac}). 
Finally, we discuss and summarise our findings in 
\S~\ref{sec:summary}.


\section{Numerical technique}
\label{sec:method}

\subsection{Simulation and Metal Cooling}
\label{sec:simulation}

We use the updated version of the Tree-particle-mesh (TreePM) smoothed 
particle hydrodynamics (SPH) code GADGET-2 \citep{Springel:05} for our 
cosmological simulations.  The gravitational dynamics is computed by a TreePM 
algorithm, which uses a particle-mesh method \citep{Hockney.Eastwood:88} for 
the long-range gravitational force and a Tree method for the short-range 
gravitational force \citep{Barnes.Hut:86}.  This hybrid algorithm makes the 
gravitational force calculation faster than a Tree method and allows better 
force resolution than a PM method in dense regions.  The gas dynamics is 
computed by an SPH method.  The SPH is particularly useful if the simulation 
needs to resolve large dynamical range, which is an inevitable requirement for
the study of galaxy formation in a cosmological context.  Therefore a 
TreePM-SPH simulation can provide a fast and high resolution calculation for 
both gravitational dynamics and hydrodynamics.

The GADGET-2 code adopts the entropy-conservative formulation \citep{Springel.Hernquist:02}, 
which alleviates the overcooling problem that previous SPH codes 
suffered from.  Our basic simulations include radiative cooling and heating 
processes for hydrogen and helium using a method similar to 
\citet*{Katz.etal:96}.  An external UV background radiation is treated as 
a spatially uniform photoionising radiation \citep{Haardt.Madau:96}, and 
modified to match the Ly$\alpha$ forest observations \citep{Dave.etal:99}.  
Implementing star formation and feedback 
from first principles is not feasible in current cosmological simulations, 
because the spatial \& mass-scales of molecular clouds are not resolved. 
Star formation and SN feedback are represented by the subgrid multiphase ISM 
model developed by \citet{Springel.Hernquist:03}. In the multiphase scheme, 
a single gas particle represents both hot and cold gas.  The stars 
are formed in the cold portion when the density exceeds a given threshold, 
$\rhoth$, which is derived self-consistently within the multiphase ISM model. 
The related parameters, such as the normalisation of gas consumption time and 
evaporation efficiency of cold gas, are set to satisfy the empirical 
Kennicutt-Schmidt law \citep{Kennicutt:98a, Kennicutt:98b}.  
The SN feedback returns some fraction of the cold gas to the hot phase, 
and increases the thermal energy of the hot gas.  

As an extension to the multiphase ISM model, the simulation includes a 
phenomenological model for SN-driven galactic wind \citep{Springel.Hernquist:03}.  
The galactic wind is particularly important for distributing the metals 
produced by SNe into the IGM.  We use the strong kinematic wind with a 
velocity of 484\,km\,s$^{-1}$.  It has been shown that this model produces
favourable results for the luminosity function of Lyman-break galaxies 
at the bright-end \citep{Nagamine.etal:04} and the H\,{\sc i} column density 
distribution function \citep*{Nagamine.etal:04-dla} at $z=3$, when compared 
to the runs without the wind. 

However, \citet{Dave.Oppenheimer:07} pointed out the problems of this 
galactic wind model by comparing with the observations of C\,{\sc iv} 
absorption lines in quasar spectra, and suggested that the momentum-driven 
wind model \citep{Murray.etal:05} is a more viable model,
which can carry more metals with lower wind velocities.  
In this paper we choose not to modify our galactic wind model in order to 
single out the effects of metal cooling, and to allow direct comparisons
to the previous works \citep{Nagamine.etal:04, Nagamine.etal:05a, 
Nagamine.etal:05b}.  Here we focus on the effects of metal cooling on galaxy 
growth, while \citet{Dave.Oppenheimer:07} focused on the C\,{\sc iv} 
statistics of the IGM.

\begin{figure*}
\centerline{\includegraphics[width=2.0\columnwidth,angle=0] {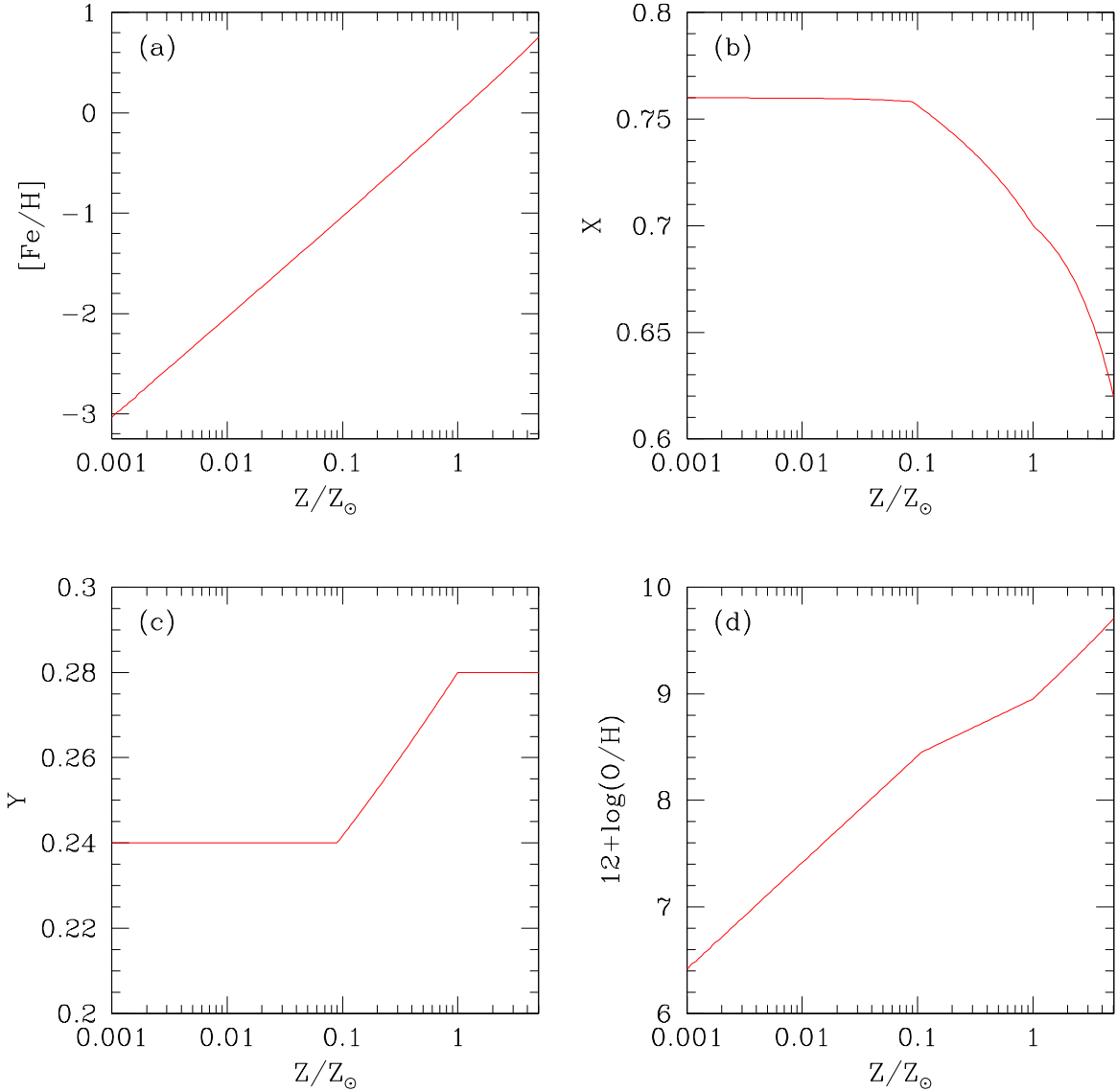}}
\caption{
The relationship between metallicity parameters in our implementation of 
metal cooling. 
The four panels show the following quantities as a function of metallicity
$Z/Z_\odot$, where $Z_\odot$ is the solar metallicity. 
[Fe/H], hydrogen mass fraction $X$, helium mass fraction $Y$, and oxygen 
abundance $12 + \log(O/H)$.  Panel (d) represents how $\alpha$-particles 
contribute to the metal evolution.
}
\label{fig:XYZ_table}
\end{figure*}

The metallicity of gas particles are also tracked by the code, assuming a 
closed box model for each gas particle. The yield 
($y = \Delta M_{\rm metal} / \Delta M_{\rm gas}$) of 0.02 is assumed. 
In principle, there could be a time delay between SN explosions and chemical 
enrichment of the ambient gas.  Unfortunately, current cosmological simulations
do not have sufficient resolution to track the detailed mixing process of 
metals.  In our simulations, we ignore this time delay and assume an 
instantaneous mixing within each gas particle.  This assumption is reasonable, 
because the overall time step of the simulations is longer than the 
mixing time-scale on small scales. 

Ideally we would like to track individual metals and compute the cooling 
rate of each element at each time-step of the simulation.  This approach has 
been used to study the evolution of individual galaxies
\citep {Recchi.etal:01,Lia.etal:02,Kawata.Gibson:03,Kobayashi:04}.
For cosmological simulations, it requires a large memory to track individual 
metals and the computation becomes exceedingly expensive 
\citep[but see][for such an attempts]{Scannapieco.etal:05,Martinez.etal:08}.
Our simulations collectively track the total mass fraction of metals 
($Z\equiv m_{\rm metal}/m_{\rm gas}$) as the measure of metallicity. 
We obtain the chemical abundance and cooling rates for a given $Z$ and
temperature by interpolating the table given in 
\citet[][hereafter SD93]{Sutherland.Dopita:93} for the standard collisional 
ionisation equilibrium model, and create a lookup table.
Metal cooling is implemented in the full range of metallicity in the
table from SD93, from [Fe/H]=-3 to [Fe/H]=1. In addition, the metal cooling 
outside of this range is computed by extrapolating the table. 
We require that the extrapolation does not go lower than the primordial cooling rate.  
The primordial abundance pattern is used where [Fe/H] $\le -1$ 
\citep{Wheeler.etal:89,Bessell.etal:91}, and the solar abundance pattern is used where 
[Fe/H] $\ge 0$ \citep{Anders.Grevesse:89}. 
For $-1 <$ [Fe/H] $ < 0$, the abundance pattern is computed by interpolating 
between the primordial and solar abundance patterns. 
Note that, [Fe/H] = $\log (n_{\textup{Fe}}/n_{\textup{H}})$ - $\log (n_{\textup{Fe}_{\sun}}/n_{\textup{H}_{\sun}})$, where $n$ is the number density.

Figure \ref{fig:XYZ_table} shows the relationships between the metal 
parameters ([Fe/H], $X$, $Y$, and $12 + \log(O/H)$) and the total 
metallicity $Z$.
Because the abundance pattern is fixed for a given $Z$, 
[Fe/H] is a monotonically increasing function of $Z$. The main difference 
between the solar and the primordial abundance pattern is the oxygen or 
$\alpha$-particle enhancement.
In the early phase of metal enrichment, Type II SNe are the main source of 
metals and they tend to generate more $\alpha$-particles than SN Ia. 
Therefore, it is expected that a low metallicity gas to have more 
$\alpha$-particles than the high metallicity one.
As mentioned above, we use two different chemical abundance pattern (at low- and 
high-end of the $Z$ values) to model this evolution in chemical composition.
We compute the oxygen abundance based on the Table 4 of SD93. 
In our code, we incorporate the metal cooling effect by adding the additional 
cooling contribution from metals on top of the primordial cooling rate, 
which is computed in a similar fashion as described in \citet{Katz.etal:96}.  

Metal enrichment changes the mean molecular weight $\mu$, which is needed to 
compute the temperature and internal energy of gas.  In order to compute 
$\mu$, we need to know the electron number density $n_e$, which varies with 
metallicity.  We obtain the values of $n_e$ from SD93 in the same way as we 
obtained the metal cooling rates.  Our electron number density estimate 
ignores the electrons from photoionised metals by the UV background 
radiation, but we expect that the number of electrons from ionised metals 
will be small. 

Varying the mean molecular weight $\mu$ also influences the SF threshold 
density $\rhoth$ in the multiphase ISM model of \citet{Springel.Hernquist:03}, 
as well as the cold and hot gas fractions of multiphase gas particles. 
In this paper, we consider two different treatments of $\rhoth$. 
In the `constant $\rhoth$' model, we take the value of $\rhoth$ as in the 
original formulation by \citet{Springel.Hernquist:03} and do not modulate
$\rhoth$ by the varying $\mu$.  
In the `varying $\rhoth$' model, we take into account of the modulation of 
$\rhoth$ by the change in $\mu$.  


\subsection{Simulation Setup}
\label{sec:IC}

We use three series of simulations with varying box size and resolution. 
The effects of metal enrichment and metal cooling are tested by comparing 
the runs listed in Table~\ref{tbl:Simulation}.
The effects of galactic wind are also tested by turning on and off the wind.
The adopted cosmological parameters of all simulations are 
$(\Omega_{m}\, , \Omega_{\Lambda}\, , \Omega_{b}\, , \sigma_{8}\, , h) \: = \: (0.3\, ,0.7\, ,0.04\, ,0.85\, ,0.7)$, which mostly concur with the results of 
the Wilkinson Microwave Anisotropy Probe (WMAP1) \citep{WMAP1}.

\begin{table*}
\centering
\begin{minipage}[t]{1.0\textwidth}
\begin{tabular}{lcccccclc}
\hline
  Run & 
  Box size \footnote{Box size in units of $h^{-1}$\,Mpc.} &  
  $N_{p}$ \footnote{Initial particle number of dark matter and gas particles.} & 
  $m_{\rm DM}$ \footnote{Dark matter particle mass in units of $h^{-1} \textup{M}_{\sun}$.}& 
  $m_{\rm gas}$ \footnote{Initial gas particle mass in units of $h^{-1} \textup{M}_{\sun}$. The star particle mass is set to $0.5 m_{\rm gas}$ in our simulations.} & 
  $\epsilon$ \footnote{Comoving gravitational softening length in units of $h^{-1}$\,kpc.  This is a proxy for the spatial resolution of the simulation, and the N216L10 series has a spatial resolution of physical $\approx 0.5$\,$h^{-1}$\,kpc at $z=3$. }&  
  $z_{\rm end}$ & 
  Metal cooling \footnote{`Primordial' means the cooling rate with primordial chemical composition.  See \S~\ref{sec:simulation} for the description of `constant $\rhoth$' and `varying $\rhoth$' models.} &
  Wind \footnote{If employed, the wind is a `strong' wind with $v_w = 484$\,km\,s$^{-1}$.}\\
\hline
 N144L10      & 10.00   & $2 \times 144^{3}$ & $2.42 \times 10^{7}$ & $3.72 \times 10^{6}$ & 2.78 & 2.75 & No (primordial) & Yes \\
 N144L10mc    & 10.00   & $2 \times 144^{3}$ & $2.42 \times 10^{7}$ & $3.72 \times 10^{6}$ & 2.78 &2.75 & Yes (const. $\rhoth$) & Yes \\
 N216L10      & 10.00   & $2 \times 216^{3}$ & $7.16 \times 10^{6}$ & $1.10 \times 10^{6}$ & 1.85 &2.75 & No (primordial) & Yes \\
 N216L10nw    & 10.00   & $2 \times 216^{3}$ & $7.16 \times 10^{6}$ & $1.10 \times 10^{6}$ & 1.85 &2.75 & No (primordial) & No \\
 N216L10mc    & 10.00   & $2 \times 216^{3}$ & $7.16 \times 10^{6}$ & $1.10 \times 10^{6}$ & 1.85 &2.75 & Yes (const. $\rhoth$) & Yes \\
 N216L10mv    & 10.00   & $2 \times 216^{3}$ & $7.16 \times 10^{6}$ & $1.10 \times 10^{6}$ & 1.85 &2.75 & Yes (varying $\rhoth$)  & Yes \\
 N288L34      & 33.75   & $2 \times 288^{3}$ & $1.16 \times 10^{8}$ & $1.79 \times 10^{7}$ & 4.69 &1.00 & No (primordial) & Yes \\
 N288L34mc    & 33.75   & $2 \times 288^{3}$ & $1.16 \times 10^{8}$ & $1.79 \times 10^{7}$ & 4.69 &1.00 & Yes (const. $\rhoth$) & Yes \\
\hline
\end{tabular}
\caption{The simulations employed in this paper}
\label{tbl:Simulation}
\end{minipage}
\end{table*}

Our naming conventions for the simulations are as follows.
The first part of the run name denotes the particle number and box size
used for the simulation series.  
The run with the extension of `mc' (for `metal cooling') adopts the 
`constant $\rhoth$' model for metal cooling, the run with the extension `mv' 
(for `metal, varying') adopts the `varying $\rhoth$' model, the run 
with the extension `nw' has no galactic wind, and the run without any 
extension uses the cooling rates for the primordial chemical composition 
(i.e., only H and He). 

The `N216L10' series is the fiducial simulation set in our study. 
We implement one simulation without galactic wind (`N216L10nw' run). 
The `N216L10' run has the same physical models as the `Q4' run used in 
\citet{Springel.Hernquist:03:S} and \citet{Nagamine.etal:04} with no metal 
cooling. 

To examine the resolution effect, we implement lower resolution simulations, 
the `N144L10' series.  Owing to the comparably small box size and missing 
long wavelength perturbations, we evolve the N144L10 and N216L10 series 
only down to $z=2.75$.  To evolve the simulation toward lower redshifts, 
we need a simulation with a larger box size. Therefore we implement the 
`N288L34' series, which has a larger box size than other series, but a 
lower resolution.  This series is evolved down to $z=1$.
The runs within the same series use the identical initial condition,
which makes the comparison more robust and enables halo-by-halo comparison 
if needed.


\subsection{Galaxy finding method}
\label{sec:Groups}

In cosmological simulations, dark matter, gas, and stars are represented 
by particles, and these particles are Monte Carlo representations of matter 
distribution.  In order to study galaxy formation using a cosmological 
simulation, we need a working definition for simulated galaxies.
In this paper the simulated galaxies are defined as isolated groups of star 
and gas particles, identified by a simplified variant of the \textup{SUBFIND} 
algorithm developed by \citet{Springel.etal:01}.  In more detail, the code 
first computes a smoothed baryonic density field to identify candidate 
galaxies with high density peaks.  The full extent of these groups are found 
by adding gas and star particles to the groups in the order of declining 
density.
If all $\Nmin$\footnote{$\Nmin$ is the minimum number of gas and star 
particles that constitute one isolated group.  In this paper we set 
$\Nmin = 32$.} nearest neighbour particles have lower densities, this group of 
particles is considered as a new group.  If there is a denser neighbour, 
the particle is attached to the group to which its nearest denser neighbour 
already belongs to.  If two nearest neighbours belong to different groups and 
one of them has less than $\Nmin$ particles, these the two groups are merged.
If two nearest neighbours belong to different groups and both of them has more 
than $\Nmin$ particles, the particles are attached to the larger group, 
leaving the other group intact.  In addition, the gas particles in groups 
should be denser than $0.01 \rhoth$. 
In this paper we are not concerned with groups without star particles. 
The fraction of such groups are very small, therefore they do not affect 
our conclusions.


\section{Global properties}
\label{sec:global}

\subsection{Cosmic star formation history}
\label{sec:cos_sfr}

\begin{figure}
\centerline{\includegraphics[width=1.0\columnwidth,angle=0] {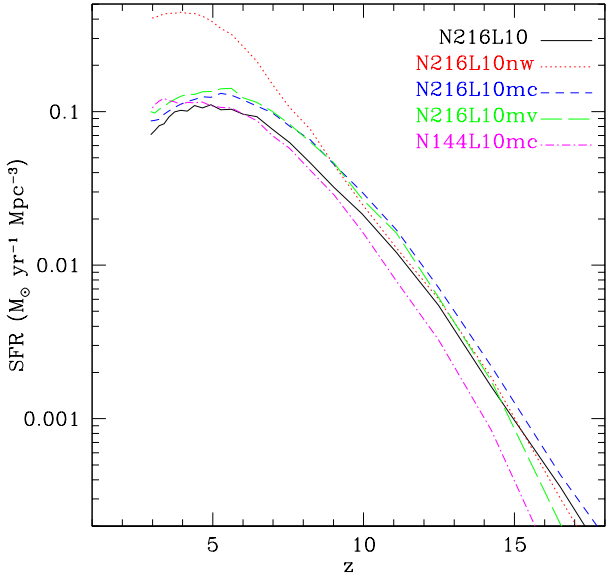}}
\caption{
Cosmic star formation histories for the N216L10 series, comparing the runs 
with and without metal cooling. 
It also shows the run without a galactic wind (N216L10nw), 
and a lower resolution run (N144L10mc). 
Overall the SFR density is increased by $\sim 20-30$\% by metal cooling
throughout the entire redshift range. 
}
\label{fig:sfr_N216L10}
\end{figure}

\begin{figure}
\centerline{\includegraphics[width=1.0\columnwidth,angle=0] {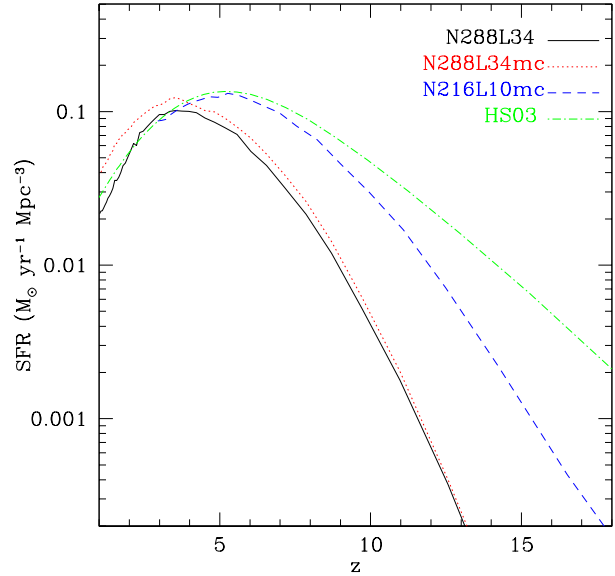}}
\caption{
The same as Figure~\ref{fig:sfr_N216L10}, but for the N288L34 series, which 
goes down to $z=1$. 
The N216L10mc run is also shown for a resolution comparison. 
The green dot-dashed line (HS03) is for the analytic model of 
\citet[][Eq.\,(2)]{Hernquist.Springel:03}. 
}
\label{fig:sfr_N288L34}
\end{figure}

Figure~\ref{fig:sfr_N216L10} shows the effect of metal cooling and galactic 
wind on the cosmic SF history for the N216L10 series.  
The comparison of N216L10mc and N216L10 runs clearly shows that metal cooling 
enhances the cosmic SFR throughout all redshifts by $20-30$\%. 
Previously, \citet[][Fig.~13]{Hernquist.Springel:03} argued that
the SFR density is enhanced by metal cooling mostly at lower redshifts 
and hardly at high-$z$, however, our simulations suggest otherwise. 
(See \S~\ref{sec:summary} for more discussion on this point.)

Metallicity needs to be higher than $\sim 10^{-2} \Zsun$ to noticeably 
increase the gas cooling rate, because the metal cooling rate is much 
smaller than that by H and He at $Z \leq 10^{-2} \Zsun$.
The metallicity of diffuse IGM is generally lower than $10^{-2} \Zsun$ at 
$z \gtrsim 3$, and the metal cooling effect on IGM is small. 
However the local SFR could be enhanced significantly even at high-$z$ when
the ISM is enriched to $Z > 10^{-2} \Zsun$, which can be easily achieved 
after a few events of SN II.  Our results suggest that star formation 
at high-$z$ can also be enhanced by metal cooling, and not just at low-$z$. 
We also find that the peak of the cosmic SF history hardly shifts by the 
introduction of metal cooling, consistently with the estimate by 
\citet{Hernquist.Springel:03}.

The N216L10mv run has a lower SFR at $z\gtrsim 9$ and a higher 
SFR at $z\lesssim 7$ than the N216L10mc run. 
This is because at high-$z$, the value of $\rhoth$ is higher in the `mv' run 
than in the `mc' run due to lower metallicity and mean molecular weight, 
leading to a larger fraction of gas not being able to satisfy the SF 
criteria in the N216L10mv run than in the N216L10mc.
The opposite is true at $z\lesssim 3$. 

Figure~\ref{fig:sfr_N216L10} also shows that the SFR in the N216L10nw run 
is higher than other runs by an order of magnitude.  This is because the gas 
can cool very efficiently without being ejected by the galactic wind. 
Although the current galactic wind models in cosmological simulations are not 
well established yet and need to be improved, both theoretical and 
observational studies evidently suggest the important role of galactic wind 
feedback \citep{Springel.Hernquist:03,Martin:05,Martin:06, Oppenheimer.Dave:06}.

As expected, the N144L10mc run has lower SFR densities at $z\ge 7$ 
compared to the N216L10 series.  This is because a lower
resolution run is unable to resolve low-mass halos that would otherwise 
forms stars in them.  When the star formation at high-$z$ is underestimated, 
some of the gas remain unconverted into stars until lower redshift, resulting
in a higher SFR at $z<4$ in the N144L10mc run than in the N216L10mc run. 

Figure~\ref{fig:sfr_N288L34} shows the cosmic SF history down to $z=1$ for 
the N288L34 series.  Again, the SFR density in the N288L34mc run is 
systematically higher than that in the N288L34 run at all redshifts.
Moreover, the difference in SFR density between the two runs increases at 
$z\lesssim 3$.  By $z \sim 3$ the IGM metallicity becomes high enough to 
noticeably increase the cooling rate, and the accretion of IGM onto galaxies 
becomes more efficient, which later fuels the star formation. 
Our results suggest that the cosmic SFR density is enhanced by metal 
cooling at $z \lesssim 3$ by both of the following two effects: 
1) more efficient IGM accretion onto galaxies due to higher metallicity of 
IGM, and 2) increased SF efficiency due to higher cooling rate
with the contribution from metal cooling. 

In Figure~\ref{fig:sfr_N288L34} the analytic model of 
\citet[][Eq.\,(2), hereafter H\&S model]{Hernquist.Springel:03} is also 
shown in the green dot-dashed line.  This model is a result of many  
cosmological SPH simulations with increasing resolution, and the authors 
gave a theoretical interpretation to the empirical fitting formula for the  
cosmic SFR density. Since our runs adopt a lower value of $\sigma_8=0.85$ 
than the original value used to derive the H\&S model ($\sigma_8=0.90$), we 
modify the H\&S model by rescaling the $\beta$ parameter in their Eq.\,(2)
and Eq.\,(47).  This reduces the SFR density at $z\gtrsim 5$ by 
$\sim 10$\%, but not much at $z\lesssim 5$. 
The results of the N288L34 and N216L10mc runs agree well with the H\&S model 
at $z<4$ and $z<6$, respectively. 
This is expected, because the N288L34 run adopts the same physical 
models as in the original simulations that were used to derive the 
H\&S model. At high-$z$, both runs fall short compared to the H\&S model 
owing to insufficient resolution.


\subsection{Phase space distribution of cosmic gas}
\label{sec:ps_gas}

\begin{figure*}
\centerline{
  \subfigure[N216L10]{\includegraphics[width=0.475\textwidth,angle=0] {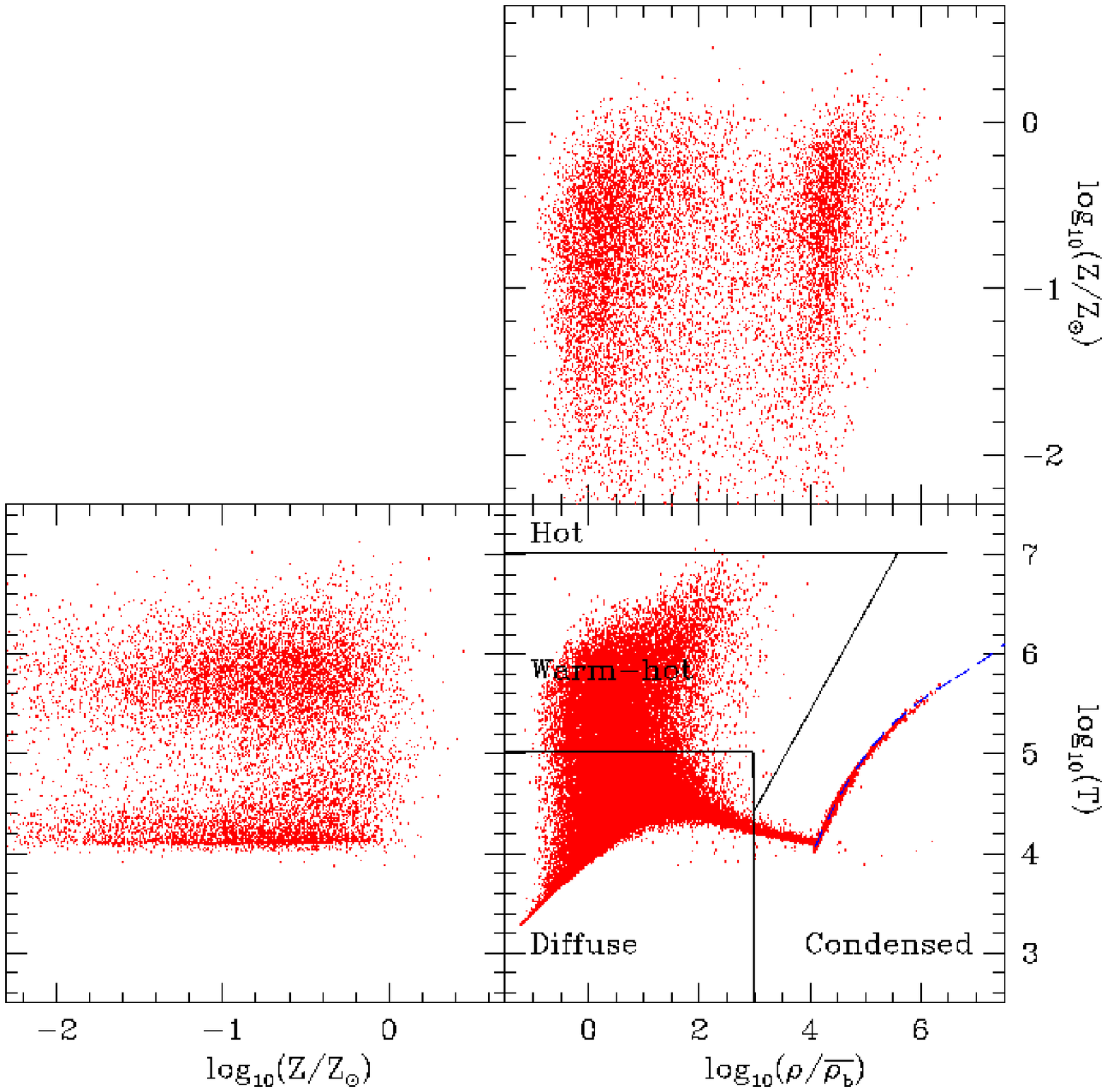}}
  \subfigure[N216L10nw]{\includegraphics[width=0.475\textwidth,angle=0] {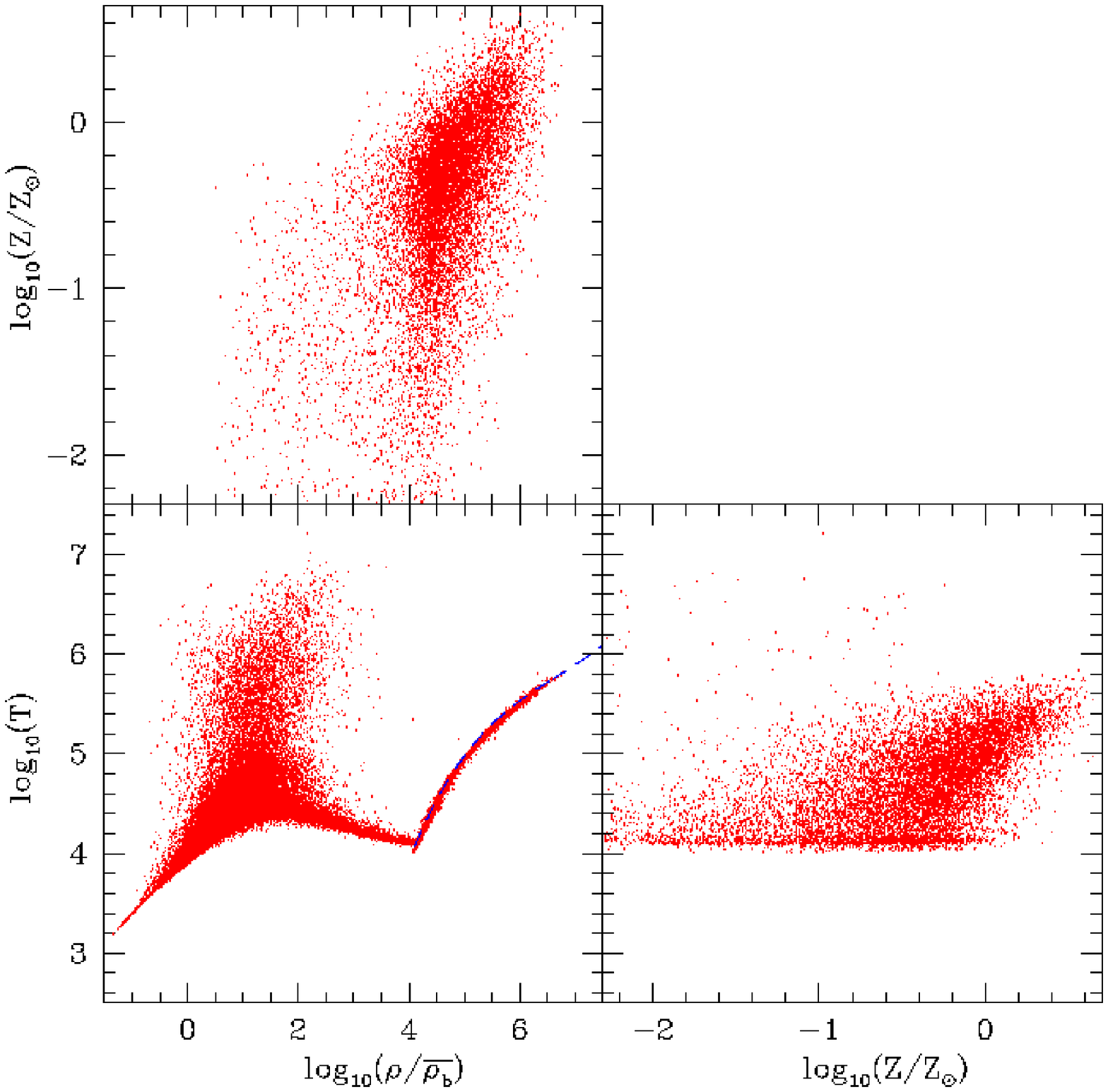}}
}\vspace{0.3cm}
\centerline{
  \subfigure[N216L10mc]{\includegraphics[width=0.475\textwidth,angle=0] {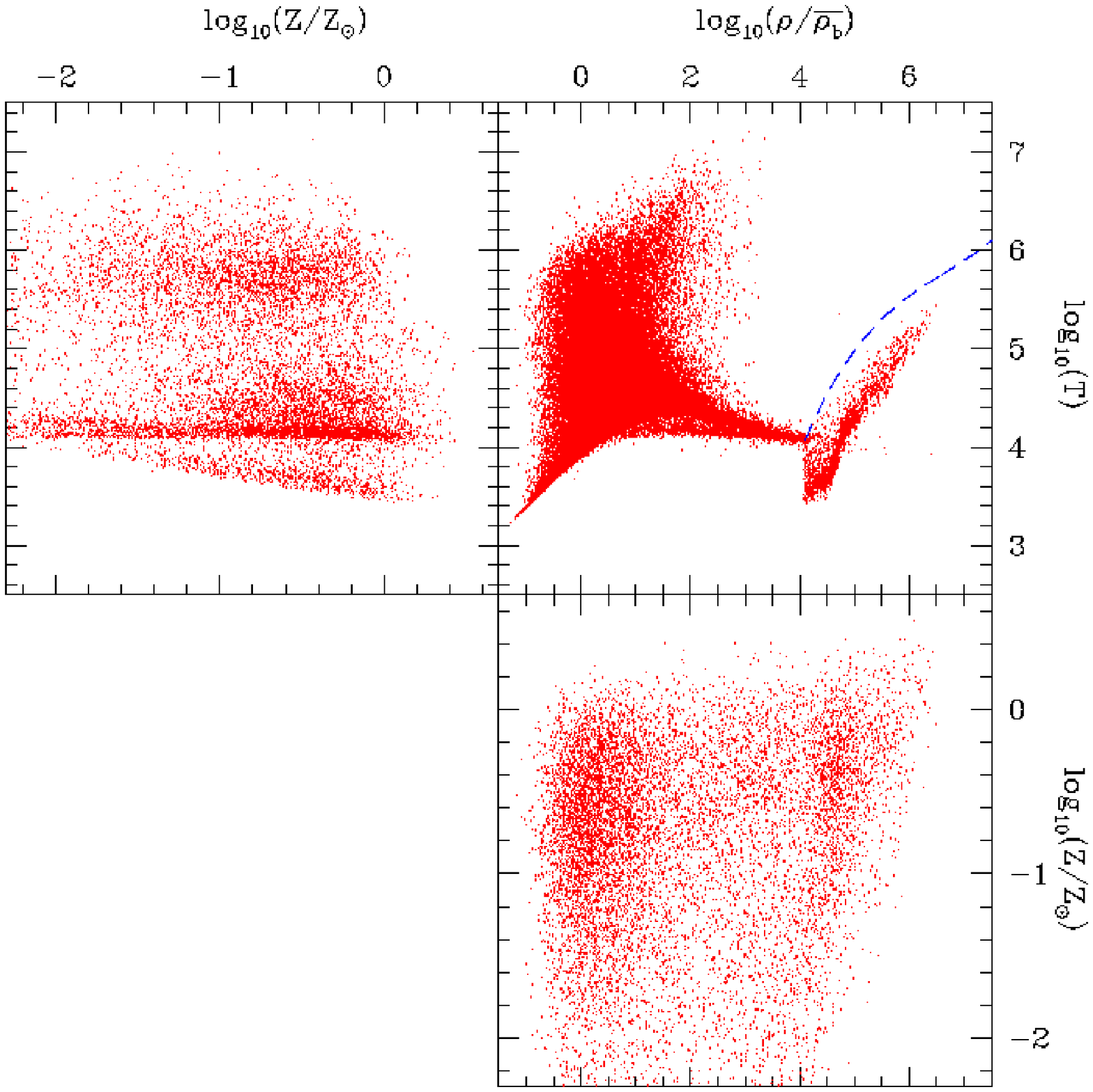}}
  \subfigure[N216L10mv]{\includegraphics[width=0.475\textwidth,angle=0] {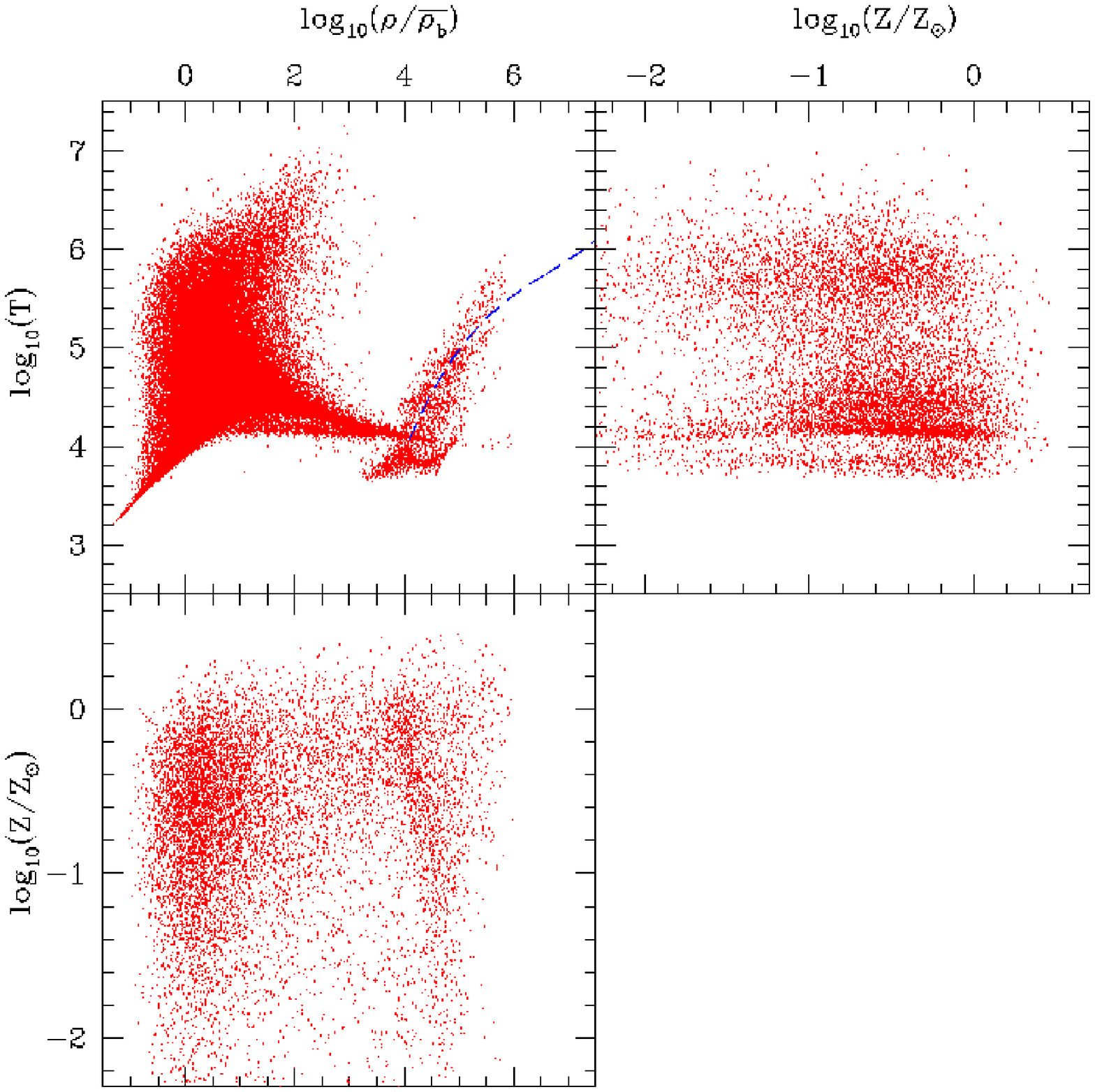}}
}
\caption{
The phase space and metallicity distribution of the cosmic gas at $z=3$ for 
the four simulations of N216L10 series.  For each run, three panels are 
shown for $\rho - T$, $Z - T$, and $\rho - Z$ relationship.  
For plotting purpose, we plot only randomly selected 1\% of the total gas 
particles in the simulation.  The blue dashed line is based on the fitting 
function to the effective equation of state for the multiphase star-forming 
gas provided by \citet{Robertson.etal:04}. 
}
\label{fig:phase.N216L10}
\end{figure*}

The metal enrichment and metal cooling also change the phase space 
distribution ($\rho - T$ diagram) of cosmic gas.  
Figure~\ref{fig:phase.N216L10} shows 
the state of cosmic gas at $z=3$ for the N216L10 series.  As shown in 
panel ($a$), the baryons in the Universe can be broadly categorised 
into four different phases according to their overdensity and temperature: 
`hot', `warm-hot', `diffuse', and `condensed' \citep{Dave.etal:99, 
Cen.Ostriker:99, Dave.etal:01}.  The first two phases are mostly the 
shock-heated gas in clusters and groups of galaxies.  
The `diffuse' phase is mostly the photoionised IGM with lower temperature, 
which can be observed as the Ly$\alpha$ forest.  The tight power-law 
relationship between $\rho$ and $T$ at $\overd \lesssim 3$ and $T<10^4$\,K 
is governed by the ionisation equilibrium \citep{Hui.Gnedin:97}, where
$\bar{\rho}$ is the mean density of baryons at $z=0$.

The `condensed' phase is the high density, cold gas in galaxies. 
Because the primordial cooling curve has a sharp cutoff at $T\simeq 10^4$\,K, 
the temperature of condensed phase decreases only down to this temperature
in Figure~\ref{fig:phase.N216L10}a,b.  The characteristic finger-like feature
at $\overd \gtrsim 10^4$ is due to the pressurisation of the star-forming gas
by SN feedback in the multiphase ISM model of \citet{Springel.Hernquist:03}. 
A fitting formula for this effective equation of state was derived by 
\citet{Robertson.etal:04}, which is shown by the blue dashed lines. 
Above the SF threshold density $\rhoth$, the gas particles in the simulation 
go into the multiphase mode and are allowed to form stars.  

The most noticeable and interesting change in the $\rho - T$ plot by the 
metal cooling is in the distribution of `condensed' gas. Star formation 
and SN explosions occur in the multiphase star-forming gas, therefore 
the immediate effect of metal cooling appears in the condensed phase. 
The presence of gas at $T < 10^4$\,K with metal cooling results 
from the effective temperature calculation for the multiphase medium,
and not from the direct gas cooling down to $T < 10^4$\,K.
According to the multiphase scheme in \citet{Springel.Hernquist:03}, 
the star forming gas consists of two phases: hot gas 
($10^5$\,K $< T <$ $10^8$\,K) and cold gas ($T = 10^3$\,K). 
The effective temperature for the multiphase gas is computed from the 
internal energy of hot and cold gases, which are derived using the cold gas
fraction. The cold gas fraction is computed from the cooling rate of 
hot gas and the gas density. The original GADGET imposes 
the effective temperature of $10^4$\,K at the SF threshold density.
In our metal cooling scheme, however, the cooling rate of hot gas increases, 
and the cold gas fraction increases. Therefore, the multiphase gas above the SF 
threshold density with higher metallicity could have lower effective 
temperature of $T < 10^4$\,K in our implementation.
 
In the N216L10mc run, the value of $\rhoth$ was kept fixed as the original 
value, which causes a sharp edge at $\overd \simeq 10^4$ for the multiphase 
gas.  In the N216L10mv run, the value of $\rhoth$ was varied 
according to the change in mean molecular weight, therefore the condensed gas 
has a spread in both temperature and density near $\overd \simeq 10^4$.

Galactic wind returns the metal-enriched, star-forming gas to the hot and 
low density IGM, as evidenced in the $\rho - Z$ diagrams. This metal-enriched, 
low-density ($\overd \lesssim 10$) gas is absent in the N216L10nw run, which 
clearly shows the necessity of galactic wind to explain the metallicity 
observed in Ly$\alpha$ forest.  
The phase distribution of IGM in the N216L10, N216L10mc, and N216L10mv runs 
are very similar.  This is presumably because the effects of metal cooling 
is not so significant yet in the IGM at $z>3$. 
As we will discuss in the next section, this is not be the case at $z<3$.


\subsection{Evolution of four phases of baryons}
\label{sec:Evolv4}

\begin{figure*}
\centerline{\includegraphics[width=2.0\columnwidth,angle=0] {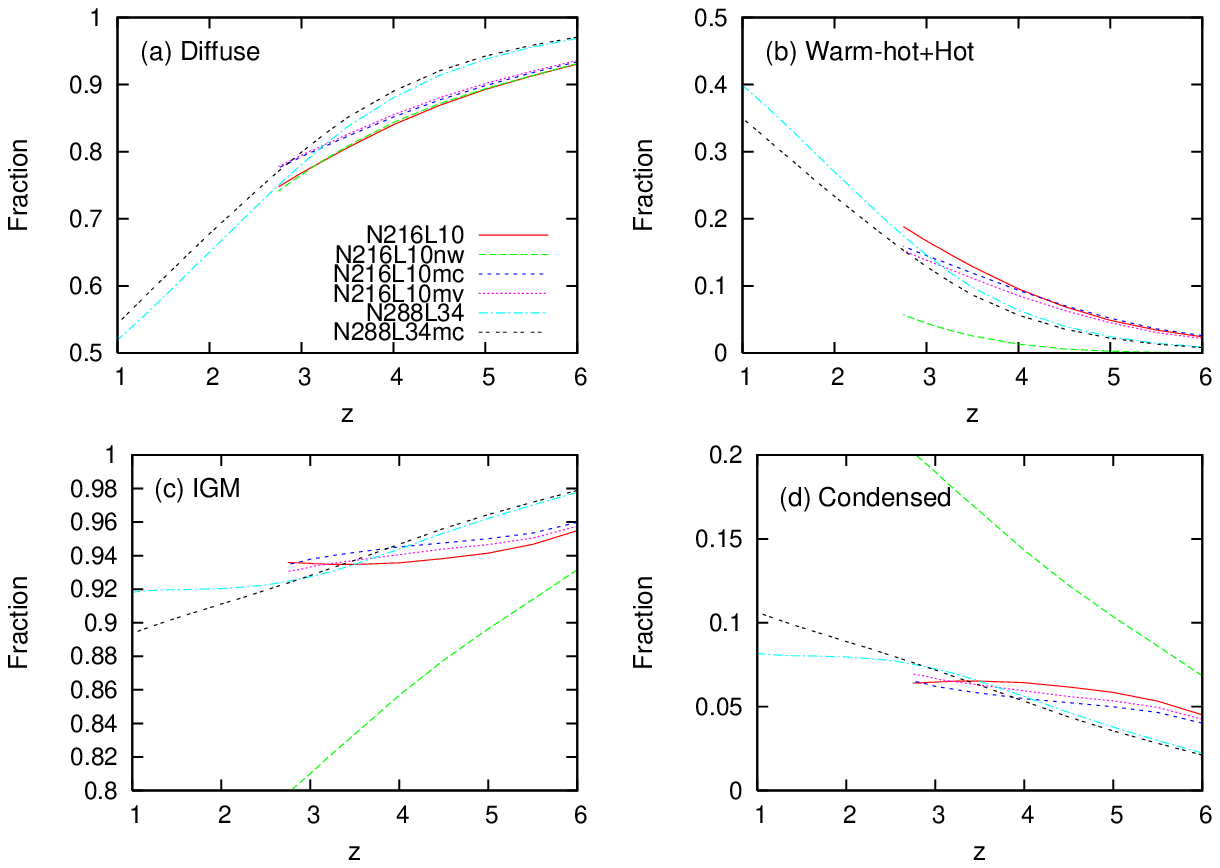}}
\caption{
The evolution of mass fractions for the four phases of baryons: hot, warm-hot, 
diffuse, and condensed.  {\it Panel (b)} shows the addition of the two 
phases, `warm-hot + hot', and {\it panel (c)} shows the total masses in the 
IGM, i.e., `diffuse + warm-hot + hot'. 
Here we compare the N216L10 and N288L34 series. 
}
\label{fig:Evolv4}
\end{figure*}

The connection between the four phases of cosmic gas is important in galaxy 
formation, and it is beneficial for us to study their evolution for a better 
understanding of metal cooling effects. 
A conventional view is that the diffuse gas is shock-heated to hot or warm-hot 
phase, which later cools down to the condensed phase \citep{Dave.etal:99, 
Cen.Ostriker:99, Dave.etal:01}.  \citet{Keres.etal:05} also suggested that 
some diffuse gas can migrate to the condensed phase without being 
shock-heated, which they called the `cold mode' accretion. 

Figure~\ref{fig:Evolv4} shows the evolution of mass fractions of different phases.
All of our simulations show that the diffuse phase continuously decreases from 
high-$z$ to low-$z$, while the `warm-hot + hot' and `condensed' phases 
continuously increase, consistently with the general expectations. 
The IGM (i.e., hot + warm-hot + diffuse) is accreted onto galaxies, 
to become the condensed phase. 
The diffuse component (panel $a$) decreases from $\approx 95$\% 
at $z=6$ to $\approx 55$\% at $z=1$.  The `warm-hot + hot' component 
(panel $b$) increases from almost zero at $z=6$ to $\approx 35$\% at $z=1$. 
The mass fraction of the total IGM (panel $c$) decreases from $96-98$\% 
at $z=6$ to $92-93$\% at $z=3$.  This reflects the increase of the  
condensed phase (panel $d$) from a few percent at $z=6$ to $\sim 7$\% at 
$z=3$.  These mass fractions are in general agreement with the results
of \cite{Dave.etal:01}.

The comparison between N288L34 and N288L34mc runs shows interesting differences. 
At $z>3$, the mass fractions of all four phases are similar in the two 
runs. However, at $z<3$, the IGM mass fraction continues to decrease in the 
N288L10mc run, while it becomes almost constant in the N288L10 run.  
A similar behaviour is seen for the condensed phase, and the N288L10mc 
continues to increase its condensed mass. 
This suggests that the metal cooling enhances the IGM accretion onto 
galaxies mostly at $z\lesssim 3$, which supplies the fuel for star formation. 
The cooling of warm+hot phase into diffuse phase is also enhanced, 
leading to a larger mass fraction of diffuse component in the N288L34mc run 
than in the N288L34 run.  (The same is true for N216L10mc and N216L10 runs.)

The N216L10nw run shows stark differences from all other runs. 
It has a significantly lower IGM mass fraction, and a very large mass fraction 
in the condensed phase.  This shows that the IGM is mostly enriched by 
galactic wind, and that most of the masses are returned into the 
`warm-hot + hot' phase, but not to the diffuse phase.


\section{Galaxy stellar \& baryonic mass function}
\label{sec:MF}

\subsection{At $z\ge 3$:}

\begin{figure*}
\centerline{\includegraphics[width=2.0\columnwidth,angle=0] {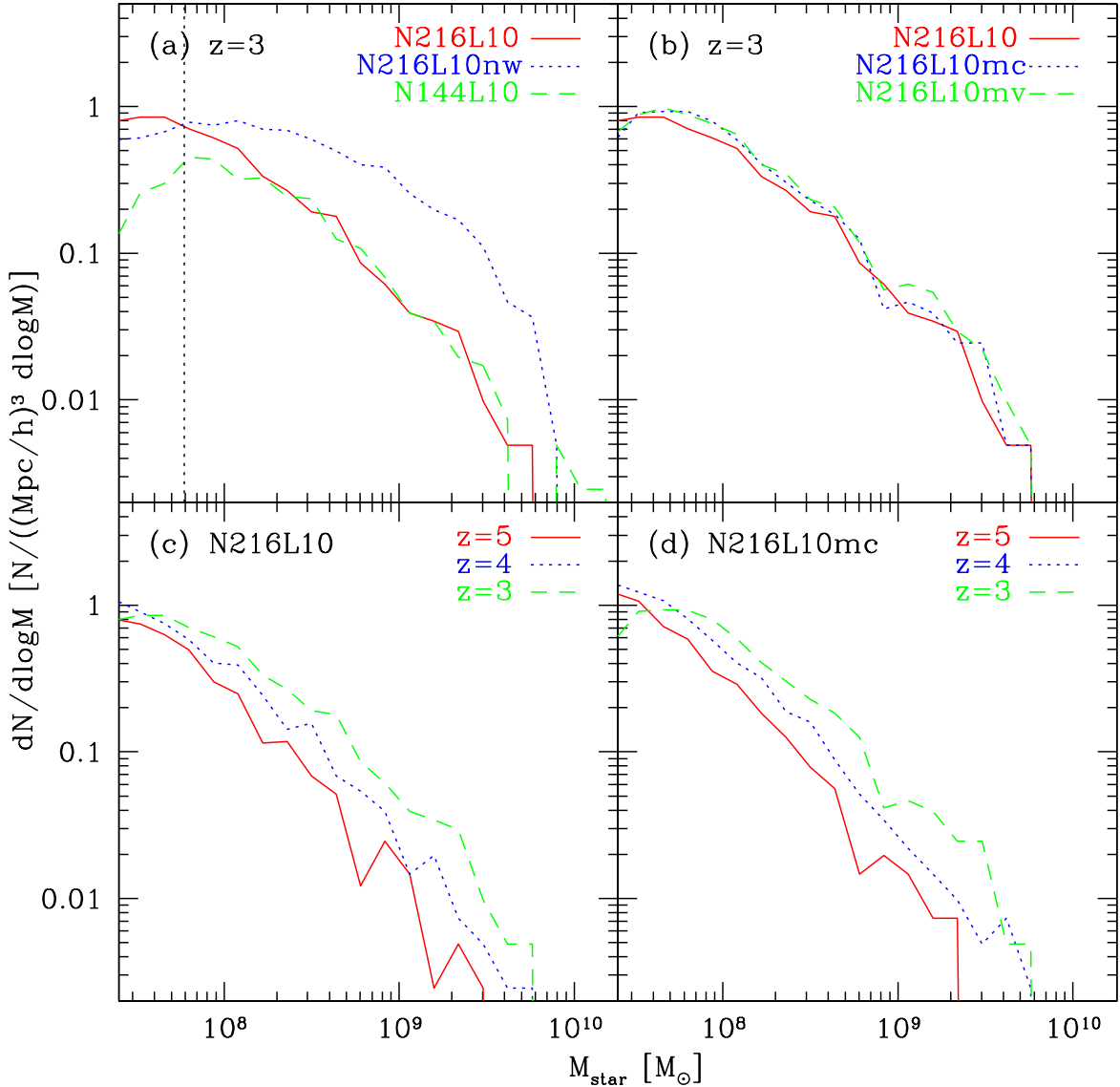}}
\caption
{
Stellar mass functions of simulated galaxies.
{\it Panel (a):} For the runs without metal cooling at $z=3$. 
The vertical dotted line indicates the galaxy masses with 32 star particles
for N144L10 run. 
{\it Panel (b)} compares the N216L10 series at $z=3$. 
The bottom two panels show the redshift evolution of GSMF in the N216L10 
({\it panel c}) and N216L10mc ({\it panel d}) run.
}
\label{fig:MF_Star}
\end{figure*}

Galaxy stellar mass function (GSMF) tells us how the stellar mass is 
distributed in different galaxies, and its redshift evolution reflects
the growth of structure in the hierarchical universe. 
Figure~\ref{fig:MF_Star} compares the GSMF in different simulations. 
We only show the range of $\Mstar \ge 2.5\times 10^7\,\Msun$, which 
corresponds to the limiting mass of 32 star particles in N216L10 runs.
We chose 32 star particles as our resolution limits because all mass 
function start to turn over around this mass range.
Panel~($a$) shows that the run with no galactic wind (N216L10nw) completely 
overestimates the GSMF, and the run with lower resolution (N144L10) 
underestimates the number of lower mass galaxies at 
$\Mstar \lesssim 10^8\,\Msun$ relative to the N216L10 run. 
Panel~($b$) shows that the metal cooling increases the masses of galaxies 
by $\sim 20$\% at $z=3$.  As a result, both the N216L10mc and N216L10mv runs 
show a slight increase in the number of galaxies at the massive end, with a 
slightly stronger enhancement in the N216L10mv run. 

Figures~\ref{fig:MF_Star}c,d demonstrate that the metal cooling also changes 
the redshift evolution of GSMF.  In the N216L10mc run, the peak of 
mass function is shifting more toward the massive side than in the N216L10 run 
at $z=3$.  From $z=4$ to $z=3$, the low-mass galaxies merge to form more 
massive galaxies, increasing the number of galaxies with 
$\Mstar > 10^8\,\Msun$ furthermore. 

 It also seems that the formation of low-mass galaxies ($\Mstar 
\lesssim 10^8\,\Msun$) is enhanced by metal cooling at early times ($z=5-6$). 
However we find that the behaviour of MF at the low-mass end is somewhat 
dependent on the threshold density for the grouping. 
As mentioned in \S~\ref{sec:Groups}, our grouping code imposes a threshold 
gas density of $0.01 \rhoth$ for a gas particle to be part of a galaxy.  
When we lower this threshold density, we find that the two MFs for 
the N216L10 and N216L10mc converged, as more particles are incorporated 
into galaxies at the outskirts of galaxies.  Metal cooling increases the 
gas density in galaxies, enabling more low-mass galaxies to satisfy the 
gas density threshold and to be identified as simulated galaxies.   
Therefore the behaviour of MF at $\Mstar \lesssim 5\times 10^7\,\Msun$ is
somewhat dependent on the choice of the outer threshold density of galaxies. 

\begin{figure*}
\centerline{\includegraphics[width=2.0\columnwidth,angle=0] {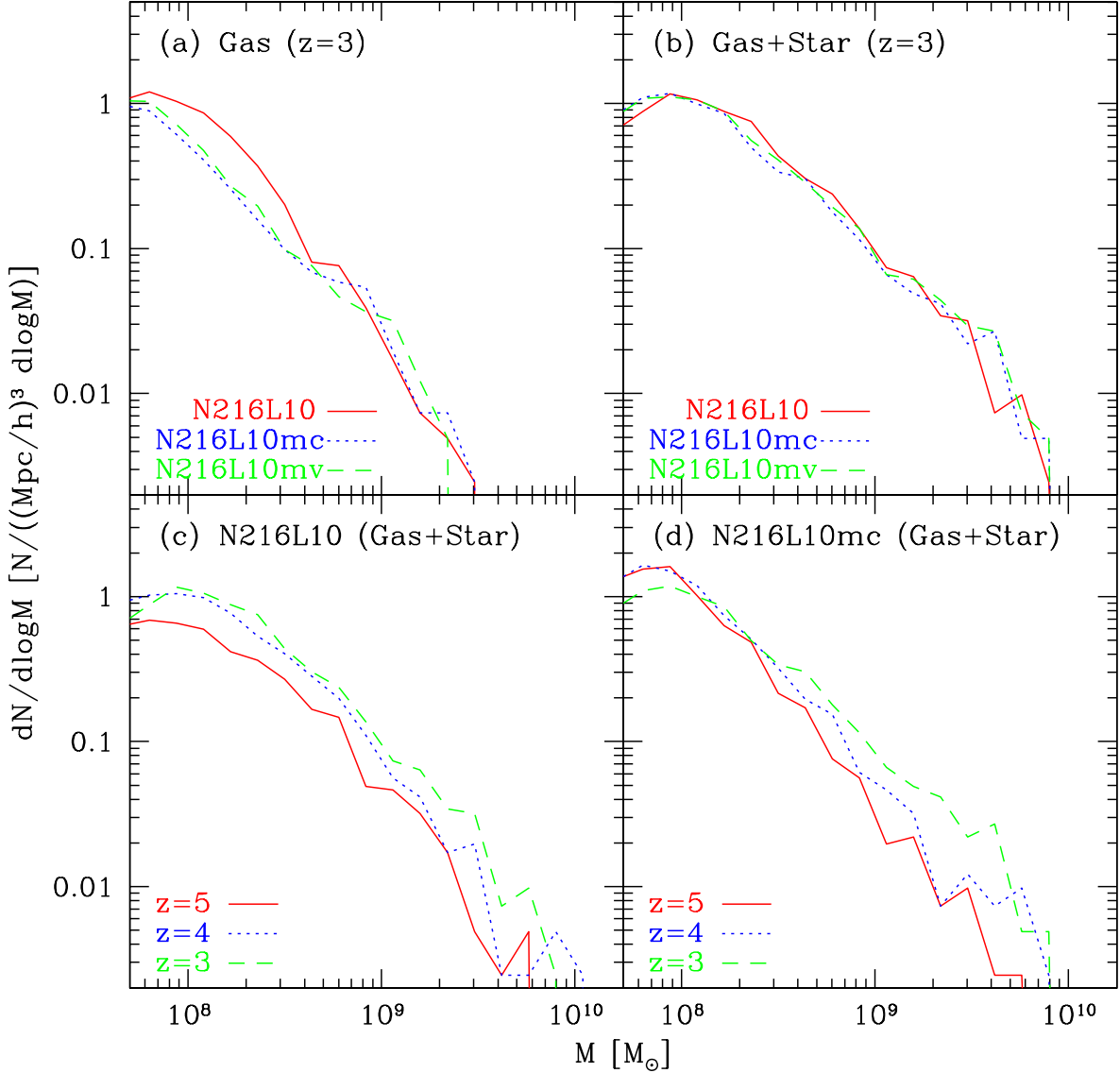}}
\caption
{
The stellar ({\it panel a}), gas ({\it panel b}), and the total baryonic 
({\it panel c}) mass functions for the N216L10 series at $z=3$. 
}
\label{fig:MF_comp}
\end{figure*}

Figure~\ref{fig:MF_comp} compares the gas and baryonic (star + gas) 
MFs for the N216L10 series at $z=3$ (panels $a$ \& $b$), and the 
redshift evolution of the baryonic MF from $z=5$ to $z=3$ in the N216L10
(panel $c$) and N216L10mc (panel $d$) runs.
Here we show only the range of $\Mstar \ge 5.0\times 10^7\,\Msun$, which 
corresponds to the limiting mass of 32 gas particles. 
Panel ($a$) shows that more gas is converted into stars in the runs with 
metal cooling (N216L10mc and N216L10mv) compared to the N216L10 run, while
panel ($b$) shows that the baryonic MF is similar in all the runs. 
As we discussed in \S~\ref{sec:Evolv4}, the IGM accretion onto galaxies is 
not so much enhanced yet before $z\sim 3$ due to relatively low IGM 
metallicity, which explains the similarity in the total baryonic MF.  
In the runs with metal cooling, the peak of GSMF is shifted 
toward higher mass, whereas the peak of gas MF is shifted toward lower mass.  
This suggests that the enhanced GSMF at $z=3$ is due to the increased 
SF efficiency by metal cooling within the galaxies. 
Again, the apparent enhancement in the number of low-mass galaxies
in the N216L10mc run compared to the N216L10 run is somewhat dependent 
on the threshold density of grouping, therefore it should be interpreted 
with caution.


\subsection{At $z< 3$:}

\begin{figure*}
\centerline{\includegraphics[width=2.0\columnwidth,angle=0] {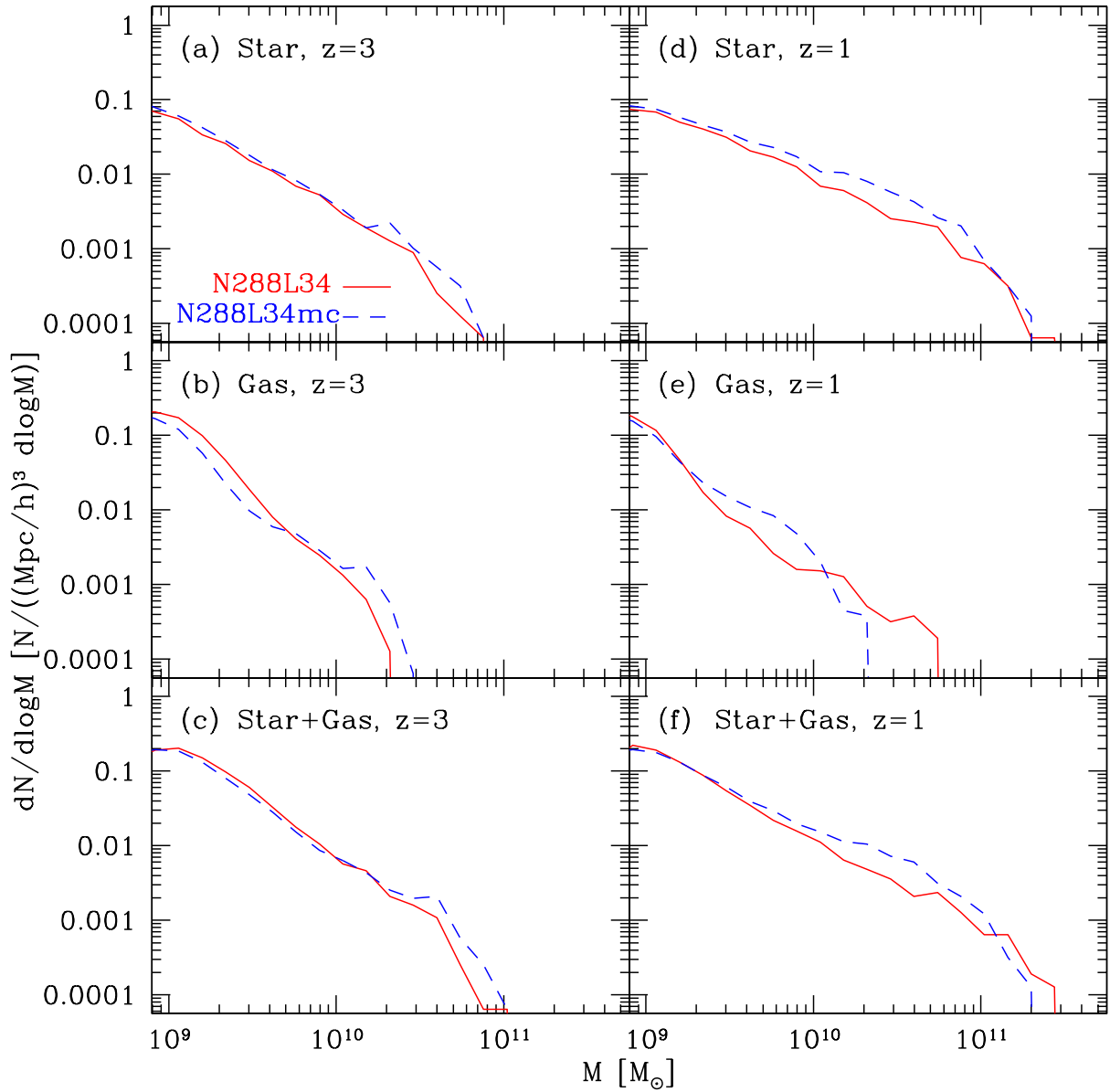}}
\caption
{
The stellar, gas, and baryonic (star + gas) mass functions at $z=3$ (left) and $z=1$ (right) for the N288L34 series. 
}
\label{fig:MF_comp_N288L34}
\end{figure*}

As the accretion of IGM onto galaxies become more efficient at $z<3$ 
due to the increased IGM cooling by the metals, we expect that the total 
baryonic mass of galaxies would be more enhanced at $z=1$ than at $z=3$.
Figure~\ref{fig:MF_comp_N288L34} shows the MFs at $z=3$ and $z=1$ 
in the N288L34 series, and it clearly demonstrates that this expectation is 
true.  Here we show only the range of $\Mgas \ge 8.2\times 10^8\,\Msun$, 
which corresponds to the limiting mass of 32 gas particles. 
The general trend in the three (star, gas, and total baryon) MFs is 
similar to that we saw in  Figure~\ref{fig:MF_comp}, although the difference 
at $z=3$ between the N288L34 and N288L34mc runs is slightly smaller than 
in the N216L10 series due to poorer resolution.
Figure~\ref{fig:MF_comp_N288L34}f shows most prominently the enhancement 
of the total baryonic MF at $z=1$ in the N288L34mc run owing to the metal 
cooling. In panels ($e$) \& ($f$), the N288L34 run actually has a longer
tail at the most massive-end than the N288L34mc run.  The reason for this 
feature is not fully clear, but it may be related to the balance between IGM 
accretion and feedback. Owing to metal cooling, IGM accretion rate increases in 
the 'mc' run, and the SFR is also enhanced, leading to a stronger
feedback. The amount of mass loss is greater in low mass galaxies,
but the net heating of IGM is more significant for massive galaxies.
The 'mc' run has stronger feedback, therefore its feedback heating may 
become more significant than IGM accretion for very massive galaxies. 
The significant IGM heating suppresses the growth of massive-end of 
mass function from $z=3$ to $z=1$ and results in shorter tail for the 'mc' run 
at $z=1$ as shown in Figure~\ref{fig:MF_comp_N288L34}e,f.


\section{Gas fraction}
\label{sec:GasFrac}

\begin{figure*}
\centerline{\includegraphics[width=2.0\columnwidth,angle=0] {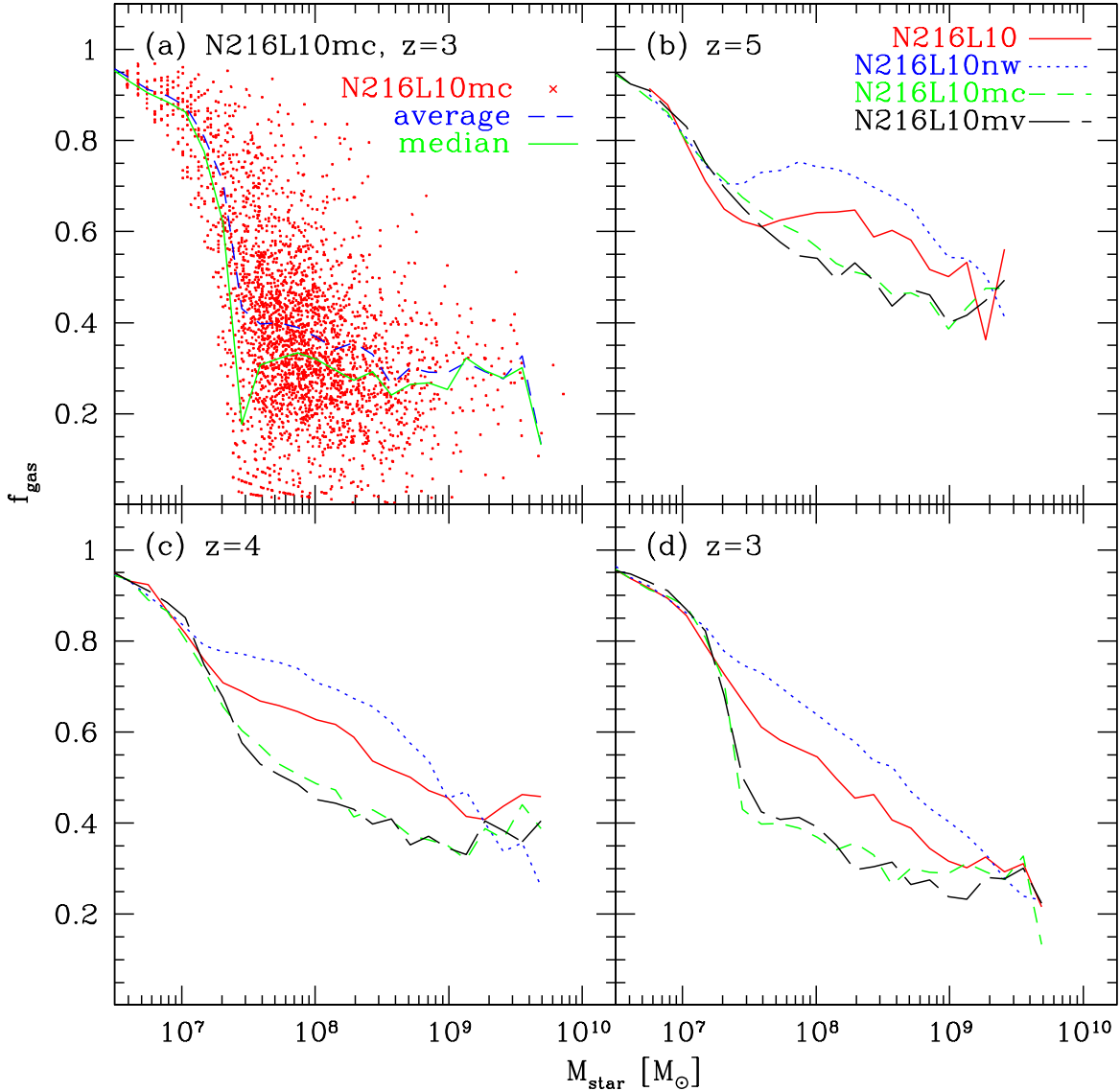}}
\caption{
The gas fraction of galaxies as a function of stellar mass.
Each data point in {\it panel (a)} is a simulated galaxy in the 
N216L10mc run at $z=3$.  Here we overplot two lines representing the 
distribution: `average' and `median' (see text for details).
Panels $(b)-(d)$ compare the gas fractions (`average' case) in the 
N216L10 series at $z=3, 4, \&\ 5$. 
}
\label{fig:GFrac_evol_N216L10}
\end{figure*}

For a fixed galaxy baryonic mass, the enhancement of star formation by metal 
cooling would decrease the gas mass fraction, 
$\fgas \equiv M_{\rm gas} / M_{\rm baryon}$.
Figure~\ref{fig:GFrac_evol_N216L10} shows $\fgas$ as a function of galaxy 
stellar mass for the N216L10 series. 
Panel ($a$) shows the data only from N216L10mc run at $z=3$, and each data 
point corresponds to a simulated galaxy. 
To characterise the distribution, we compute the following two quantities 
in each logarithmic stellar mass bin: `average' and `median'.
The `average' is the ratio of total gas mass to total baryonic mass for 
all the galaxies in each mass bin, i.e., 
$\sum_{i}^{} M_{\rm gas, i} / \sum_{i}^{} M_{\rm baryon, i}$. 
The `median' case is simply the median of $\fgas$ values in each mass bin.
Both quantities show a similar trend, however, there is a sharper drop-off at 
$\Mstar \simeq 2.5 \times 10^{7}\,\Msun$ for the `median' case in 
Figure~\ref{fig:GFrac_evol_N216L10}a.  This mass-scale corresponds to 
32 star particles in the N216L10 series.  
We find that there are many galaxies with $\fgas = 0$ above this 
mass-scale, which causes the sharp drop-off in the `median' line. 
Below this limiting mass, galaxies are not resolved well, which results in 
an underestimate of star formation and an overestimate of $\fgas$. 
If we had a higher resolution simulation with finer particle masses, this 
limiting mass-scale would shift to a lower mass.  Therefore the location of 
this sharp drop-off is currently determined by the resolution of our 
simulation.  
However, dark matter halos would stop forming stars at some 
lower limiting halo mass, if we had an infinitely high-resolution simulation, 
This lower limit to the galaxy mass is presumably determined by the 
photoevaporation of gas by the UV background radiation \citep{Rees:86, 
Efstathiou:92, Quinn.etal:96,Gnedin:00, Nagamine.etal:04-dla, Pontzen.etal:08,
Okamoto.etal:08}.  Recent works suggest that star formation 
could be suppressed by the UV background in halos with 
$M_{\rm halo} \lesssim 10^9\,\Msun$ at $z\sim 3$.  Galaxies with 
$\Mstar \simeq 2 \times 10^{7}\,\Msun$ would reside in halos with 
$M_{\rm halo} \approx 2\times 10^9\,\Msun$. 
The N216L10 series would resolve such a halo with $\sim 280$ dark matter 
particles, and its mass resolution is actually close to the 
astrophysical limit for dwarf galaxy formation at $z\sim 3$.  
Therefore the sharp drop-off in $\fgas$ at $\Mstar \simeq 2 \times 10^{7}\,\Msun$ 
may not be so far from the true answer.

\begin{figure*}
\begin{center}
\includegraphics[width=1.0\columnwidth,angle=0] {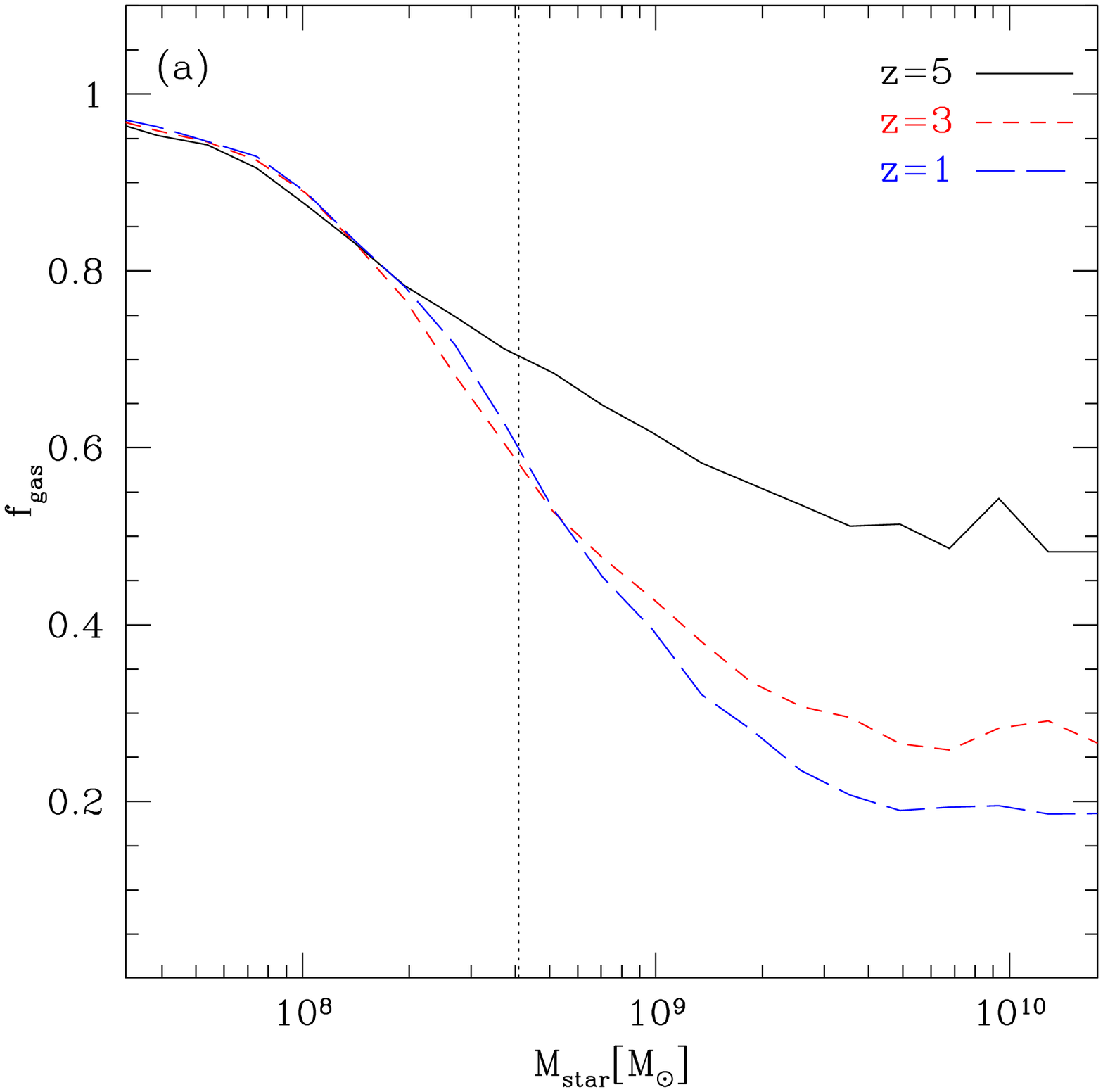}
\includegraphics[width=1.0\columnwidth,angle=0] {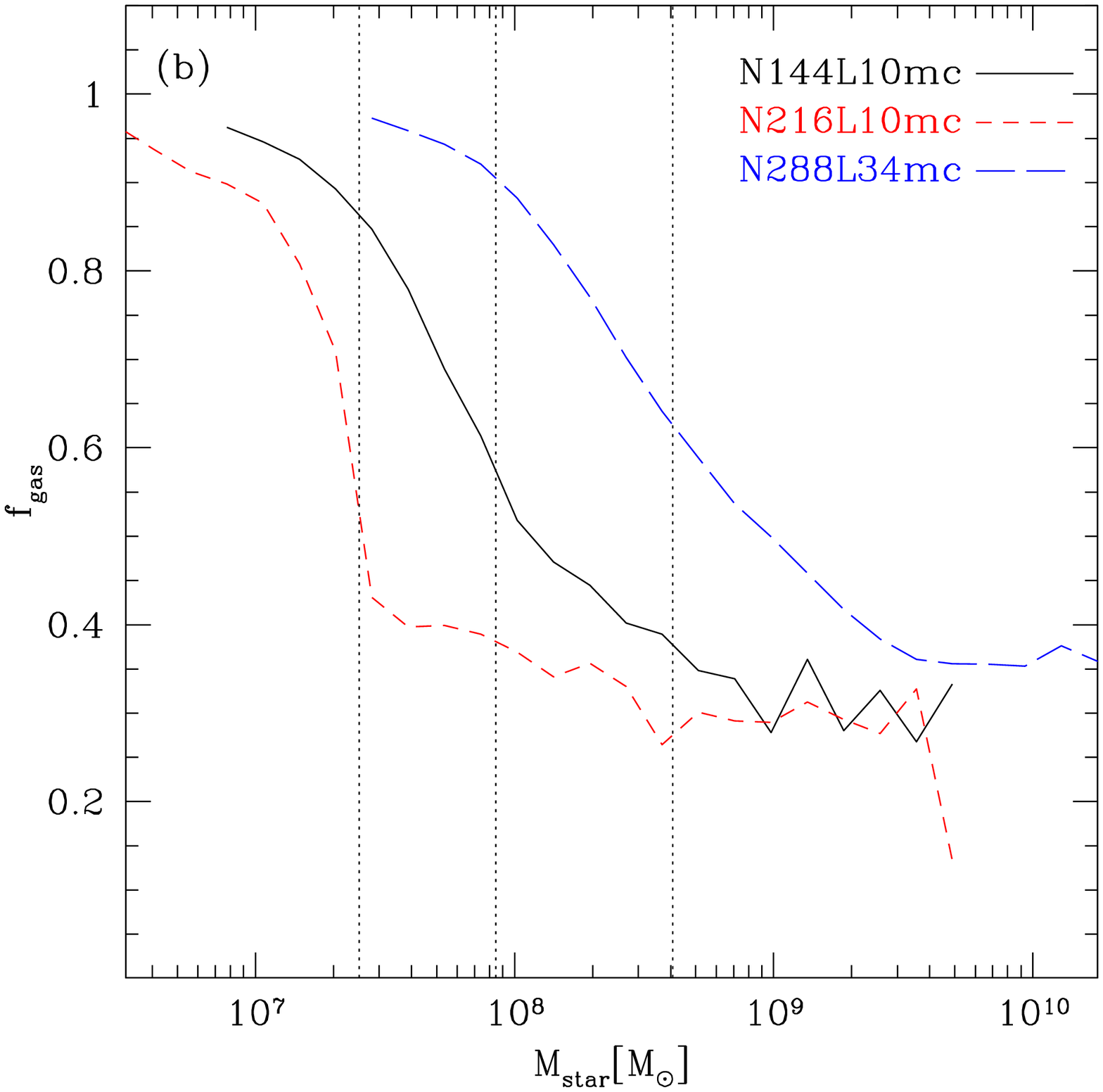}
\caption{
{\it Panel (a)} highlights the redshift evolution of $\fgas$ in the 
N288L34 run from $z=5$ to $z=1$.  
{\it Panel (b)} shows the effect of numerical resolution on $\fgas$. 
The vertical dotted lines indicate the galaxy masses with 32 star particles
for each run.  The simulation results shift to lower masses as the resolution 
is increased. We plot the `average' case in these figures. 
}
\label{fig:MGFrac_resolution}
\end{center}
\end{figure*}

We find that $f_{gas}$ increases with decreasing $\Mstar$ at all redshifts, 
regardless of metal cooling and wind effects.
This trend is qualitatively consistent with current observations.
\citet{Erb.etal:2006} estimate the gas fraction as a function of stellar mass
using the rest-frame UV-selected star-forming galaxies at $z \sim 2$, and show
that the gas fraction decreases with increasing stellar mass. And using 
the mean gas and stellar mass, they find the average $f_{gas} \sim 0.35 $, 
which agrees with the predicted gas fraction for massive galaxies in our simulation.
In addition, \citet{Geha.etal:06} reported that the average neutral 
gas fraction is $\langle \fgas \rangle = 0.6$ for the local dwarf
galaxies selected from the Sloan Digital Sky Survey.  In our N216L10 series
with metal cooling, the `average' $\fgas$ reaches 0.6 for galaxies with
$\Mstar \simeq 2\times 10^7\,\Msun$ at $z=3$ \& 4. 

Figures\,\ref{fig:GFrac_evol_N216L10}b,c,d show the redshift evolution of 
$\fgas$ for the N216L10 series.  In the runs with metal cooling (N216L10mc and 
N216L10mv), $\fgas$ is lower than in the N216L10 run by $20-30$\% at all 
redshifts for galaxies with $\Mstar = 10^{7.5} - 10^9\,\Msun$.  
This result suggests that the metal cooling reduces $\fgas$ owing
to more efficient star formation. 
The values of $\fgas$ seem to be more convergent at the massive-end 
($\Mstar > 10^9\,\Msun$). 
In addition, we find that galaxies in the N216L10nw run are the most gas-rich, 
because almost no galactic gas is returned to the IGM.

Figures~\ref{fig:GFrac_evol_N216L10}b,c,d also show that $\fgas$ is higher 
at higher redshifts.  For example, the well-resolved galaxies 
with $\Mstar = 10^9\,\Msun$ have $\fgas \simeq 0.45$ at $z=5$, but 
$\fgas = 0.25$ at $z=3$.  As we saw in Figure~\ref{fig:MF_Star}d, there is a
considerable growth in GSMF from $z=5$ to $z=3$.  Together with the decrease
in the gas fraction, these results suggest that most of the gas accreted 
during $z=3-5$ has been converted into stars. 

Figure~\ref{fig:MGFrac_resolution}a highlights the redshift evolution of 
$\fgas$ from $z=5$ to $z=1$ in the N288L34mc run. The gas mass fraction 
clearly decreases with decreasing redshift as the gas is converted into 
stars.  The rate of decrease is greater at $z=5 \rightarrow 3$ than at 
$z=3 \rightarrow 1$, with $\fgas \simeq 0.5, 0.25,$ \& 0.2 for $z=5, 3$, 
\& 1, respectively, for galaxies with $\Mstar \gtrsim 3\times 10^9\,\Msun$.

Figure~\ref{fig:MGFrac_resolution}b shows the effect of resolution on $\fgas$. 
The location of the drop-off in $\fgas$ 
shifts to lower masses as the resolution is increased from N144L10mc to 
N216L10mc run.  The location of the drop-off is at around our resolution limit --
the vertical lines in {\it Panel (b)}. Therefore the values of $\fgas$ are 
overestimated in unresolved galaxies.


\section{Discussions and Conclusions}
\label{sec:summary}

Using cosmological hydrodynamic simulations with metal enrichment and metal 
cooling, we studied their effects on galaxy growth and cosmic SFR.  
Owing to metal cooling, the SFR density increases about 20\% 
at $z=3$ and about 50\% at $z=1$.
Our results suggest that metal cooling enhances the star formation 
through two different processes: 
1) more efficient conversion of local ISM into stars (i.e., increase of  
local SF efficiency), and 
2) the increase of IGM accretion onto galaxies. 
The former process is in effect essentially at all times at $z\lesssim 15$, 
because the local ISM can be instantaneously enriched by SN explosions.
This process enhances the SFR, but does not noticeably change the total 
baryonic mass of galaxies. 
The latter process, on the other hand, can increase the total baryonic mass
of galaxies, as well as enhancing the overall SFR density by supplying more 
gas for star formation.   This process becomes effective only at lower 
redshifts ($z\lesssim 3$), because it takes some time to enrich the 
IGM up to $Z \gtrsim 10^{-2}\Zsun$.  

In this paper, we used two different prescriptions for star formation 
in the runs with metal cooling:  
1) the `constant $\rhoth$' scheme and 2) the `varying $\rhoth$' scheme. 
The value of $\rhoth$ was fixed in the former scheme, while 
it was modulated according to the metallicity in the latter scheme. 
Both schemes show similar increase of cosmic SFR density (see 
Figure~\ref{fig:sfr_N216L10}), which implies that the overall SF enhancement 
by metal cooling is not so sensitive to the choice of $\rhoth$
(but with some differences as we discussed in \S~\ref{sec:cos_sfr}).

\begin{figure*}
\begin{center}
\includegraphics[width=1.9\columnwidth,angle=0] {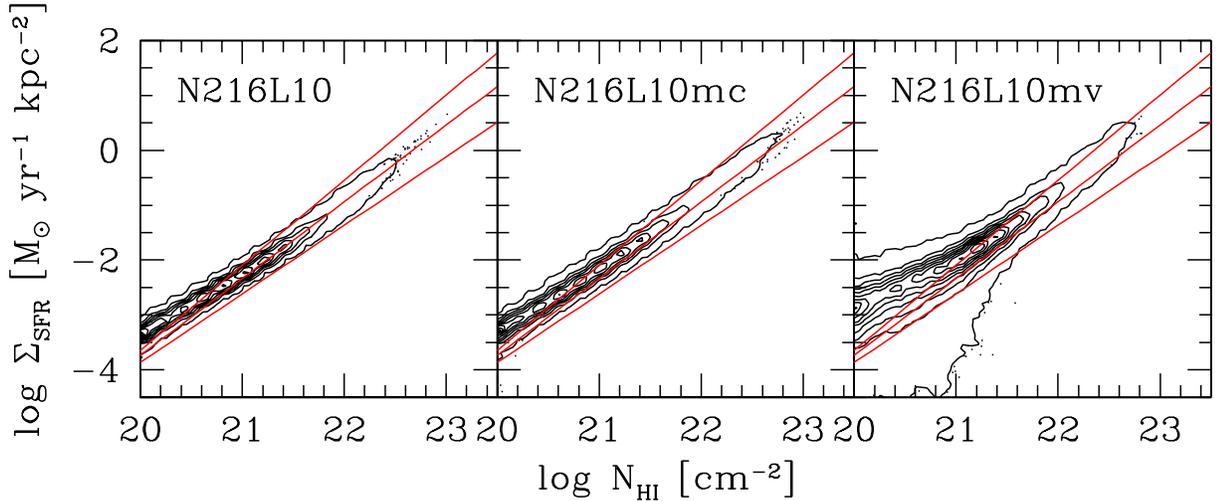}
\caption{Comparison of the N216L10 series with the empirical Kennicutt law. 
\label{fig:kenni}}
\end{center}
\end{figure*}

The multiphase ISM model for star formation by \citet{Springel.Hernquist:03} 
contains two free parameters: $\rhoth$ and the normalisation of gas 
consumption time-scale, $t_\star^0$.  The values of these two parameters were 
originally chosen to match the empirical Kennicutt-Schmidt law 
\citep{Kennicutt:98a,Kennicutt:98b} using simulations of isolated disk 
galaxies.  Since we did not change the value of $t_\star^0$ in the runs
with metal cooling (as well as $\rhoth$ in the `mc' runs), 
it is possible that our simulations may now violate the Kennicutt law.  

However, as we show in Figure~\ref{fig:kenni}, the plots of $\NHI$ vs. 
$\Sigma_{\rm SFR}$ for the N216L10 and N216L10mc runs are not so different 
from each other. 
This can be understood as follows.  As Figure~\ref{fig:Evolv4}c showed,  
metal cooling increases the density and lowers the temperature of 
star-forming gas.  The fundamental scaling relationship between SFR and 
cold gas density does not need to change when the metal cooling is introduced; 
the star-forming gas particles would simply slide upward along the 
Kennicutt law, $\Sigma_{\rm SFR} \propto \NHI^{1.4}$. 

In the case of N216L10mv run, we varied $\rhoth$ according to the gas 
metallicity, primarily scattering it to lower values because higher metallicity
increases the cooling rate and lowers $\rhoth$.  Therefore more gas particles 
become eligible to form stars and $\Sigma_{\rm SFR}$ is scattered upward 
above the Kennicutt law for a given value of $\NHI$, resulting in 
a broader deviation from the power-law relationship of the Kennicutt law near
$\log \NHI \simeq 20$.  At higher values of $\NHI$, $\Sigma_{\rm SFR}$ is
enhanced, but still within the range of the Kennicutt law. 
This difference can be alleviated by adjusting the SF timescale 
\citep[see][]{Springel.Hernquist:03}, however, we try to emphasise the effect
of metal cooing on SFR by keeping this parameter the same in all the runs.
Further investigations of the \HI\ aspect of our simulations is beyond the 
scope of the present paper, and we will present the results elsewhere. 

Upon evaluating the effect of metal cooling, \citet{Springel.Hernquist:03:S} 
considered two different gas phases where the cooling become important. 
One is the diffuse gas in galactic halos, which must radiate its thermal 
energy in order to collapse onto the high-density ISM, and the other is the 
multiphase star-forming gas.  They argued in their \S~5.2 that the cosmic SFR 
density at $z\gtrsim 6$ would not be affected by the metal cooling very much, 
because at high-$z$ cooling is so efficient that the gas in diffuse 
atmospheres of halos cools nearly instantly, even without any metal cooling.  
They expected their model results to be largely independent of metal 
enrichment, because in this regime the evolution of SFR density is driven by 
the fast gravitational growth of the halo mass function. 
For the star-forming multiphase ISM, they argued that the parameters of 
their model (e.g. cold gas evaporation efficiency and gas consumption 
time-scale) can be adjusted such that the normalisation of the Kennicutt law 
can be maintained, yielding to first order unaltered model predictions. 
However, our simulations, in particular the comparison between the N216L10 and 
N216L10mc run, suggest that the metal cooling can enhance the SFR density 
even at $z\gtrsim 6$ by $\sim 20-30$\%.  This is because the gas density 
becomes higher and temperature becomes lower (see Figure~\ref{fig:Evolv4}c) 
by responding to the enhanced cooling rate by metals. 

In summary, we consider that our study serves to demonstrate the importance 
of metal enrichment and metal cooling in galaxy formation, beyond 
our original motivation to perform a simple numerical study. 
Our study highlights the role of metals at low-$z$ and high-$z$ to 
enhance the star formation in the Universe, and demonstrates that the metals
increase the SF efficiency in the star-forming ISM, as well as 
enabling more IGM to accrete onto galaxies and fuel star formation.
We expect that metal cooling would also alter some of the galaxy properties 
such as the mass-metallicity relationship \citep[e.g.,][]{Tremonti.etal:04} 
and specific SFR \citep[e.g.,][]{Noeske.etal:07}.  We plan to investigate 
these issues in the future, as well as alternative models for star formation
and feedback.


\section*{Acknowledgements}
We thank V. Springel for allowing us to use the updated version 
of {\small GADGET-2} code for our study, and for useful comments on 
the manuscript. 
KN is grateful for the hospitality of Institute for the Physics and 
Mathematics of the Universe, University of Tokyo, where part of this work was done.
We are also grateful to the anonymous referee for constructive comments which improved this paper.
This research was supported in part by the National Aeronautics and 
Space Administration under Grant/Cooperative Agreement No. NNX08AE57A 
issued by the Nevada NASA EPSCoR program, by the National Science 
Foundation through TeraGrid resources provided by the San Diego 
Supercomputer Center (SDSC), and  by the President's Infrastructure Award at UNLV. 
The simulations were performed at the UNLV Cosmology Computing Cluster
and the Datastar at SDSC.


\end{document}